%% file: main.tex
\documentclass[letter,10pt, conference]{ieeeconf}
\IEEEoverridecommandlockouts
\usepackage{amssymb,amsmath,color,subfigure}
\usepackage{graphicx}
\usepackage{xspace,stackrel,hyperref}
\usepackage{wrapfig}
\usepackage{mathtools}
\usepackage{tikz,pgfplots}
\usepackage{etoolbox}
\usetikzlibrary{external} 
\usepackage{autonum}
\usepackage{graphicx}

\definecolor{color0}{rgb}{0.847058823529412,0.749019607843137,0.847058823529412}
\definecolor{color1}{rgb}{0.576470588235294,0.43921568627451,0.858823529411765}
\definecolor{color2}{rgb}{0.294117647058824,0,0.509803921568627}

\usetikzlibrary{shapes,positioning}

\usepackage{dsfont}

\definecolor{mycolor4}{RGB}{230,97,1}
\definecolor{mycolor2}{RGB}{178,171,210}
\definecolor{mycolor3}{RGB}{253,184,99}
\definecolor{mycolor1}{RGB}{94,60,153}

\makeatletter 
\pretocmd\@bibitem{\color{black}\csname keycolor#1\endcsname}{}{\fail}
\newcommand\citecolor[1]{\@namedef{keycolor#1}{\color{blue}}}
\makeatother

\DeclareMathAlphabet{\mathcal}{OMS}{cmsy}{m}{n}

\def\beq{\begin{equation}}
\def\eeq{\end{equation}}

\newcommand{\mc}{\mathcal}

\newcommand{\ul}{\underline}

\newcommand{\R}{\mathds{R}}

\renewcommand{\d}{\text{d}}

\DeclareMathOperator*{\argmin}{arg\,min}

\definecolor{mycolor1}{RGB}{230,97,1}
\definecolor{mycolor2}{RGB}{178,171,210}
\definecolor{mycolor3}{RGB}{253,184,99}
\definecolor{mycolor4}{RGB}{94,60,153}
\definecolor{mycolor5}{rgb}{0,0,0}

\newtheorem{theorem}{Theorem}
\newtheorem{proposition}{Proposition}

\newtheorem{remark}{Remark}

\newtheorem{example}{Example}

\usepackage{subfigure}
\newcommand{\subfigref}[2]{\hyperref[#1]{\ref{#1}(#2)}}

\newtoggle{full_version}
\toggletrue{full_version}

\title{\LARGE\bf  Macroscopic pricing schemes for the utilization of pool ride-hailing vehicles in bus lanes}
\author{Lynn Fayed, Gustav Nilsson, and Nikolas Geroliminis \thanks{
L. Fayed, G. Nilsson, and N. Geroliminis are with the School of Architecture, Civil and Environmental Engineering, École Polytechnique Fédérale de Lausanne (EPFL), 1015 Lausanne, Switzerland. {\tt\small \{lynn.fayed, gustav.nilsson, nikolas.geroliminis\}@epfl.ch}. This work was supported by the Swiss National Science Foundation under NCCR Automation, grant agreement 51NF40\_180545.}\iftoggle{full_version}{}{\thanks{An extended version containing all the proofs is available at \url{https://arxiv.org/abs/XXXXX}}
}}

\begin{document}

\maketitle
\thispagestyle{empty}
\pagestyle{empty}

\begin{abstract}
With the increasing popularity of ride-hailing services, new modes of transportation are having a significant impact on the overall performance of transportation networks. As a result, there is a need to ensure that both the various transportation alternatives and the spatial network resources are used efficiently. In this work, we analyze a network configuration where part of the urban transportation network is devoted to dedicated bus lanes. Apart from buses, we let pool ride-hailing trips use the dedicated bus lanes which, contingent upon the demand for the remaining modes, may result in faster trips for users opting for the pooling alternative. Under an aggregated modelling framework, we characterize the spatial configuration and the multi-modal demand split for which this strategy achieves a system optimum. For these specific scenarios, we compute the equilibrium when ride-hailing users can choose between solo and pool services, and we provide a pricing scheme for mitigating the gap between total user delays of the system optimum and user equilibrium solutions, when needed.
\end{abstract}

\section{Introduction}

The flexibility and convenience of ride-hailing services are key features of this on-demand transportation alternative. This is because they offer door-to-door rides, and they are characterized by low fares and short waiting times. However, soon after their launch, many problems have surfaced regarding their operation, safety, and impact on traffic. In particular, the high number of idle ride-hailing vehicles, while providing a good level of service, significantly deteriorates traffic conditions as shown in~\cite{inefficiency_2021_beojone}. Ride-splitting, where passengers pool their rides with other users, can mitigate this negative impact because it allows ride-hailing drivers to travel for shorter distances while serving the same demand. Pooling passengers receive a fare discount to compensate for the longer travel time they incur but the pool engagement levels are still moderate. 

To improve the overall understanding of the operation and impact of ride-splitting, many researchers have focused on modelling ride-hailing services to i) explore the pricing mechanism of these markets and ii) quantify their impact on traffic and other transportation modes. Through a comprehensive modelling of ride-hailing supply and demand in~\cite{surge_2017_castillo}, the authors highlighted the importance of surge pricing in the event of a supply shortage. In~\cite{ridepooling_2020_ke}, the authors computed the maximum achievable solo and pool demand that can be serviced by the network and proved that pooling is able to reduce the network travel time for pool riders and private vehicles concurrently utilizing the same network. 

Since the dynamics in urban transportation networks are highly complex, it is necessary to formulate tractable models suitable for theoretical analysis. One such approach is the use of network-level Macroscopic Fundamental Diagrams (MFDs), which represent an aggregate relationship between traffic flow, density, and speed in a region~\cite{geroliminis2008exsistence}. Their potential to provide an accurate estimate of aggregate traffic measures is observed in multiple locations as shown in~\cite{emperical_2017_menendez}. This paved the way for the use of MFD modelling for many different applications, such as multi-region perimeter control~\cite{optimal_2013_geroliminis} which goal is to reduce traffic congestion in urban areas. 
Beyond macroscopic modelling of car traffic dynamics, MFDs are very versatile as they can also capture the interactions between vehicle and bus traffic, as in~\cite{three_2014_geroliminis} where the authors used 3D-MFD to show that the marginal influence on traffic of buses is not equivalent to that of cars.
This latter tool is used in~\cite{FAYED202329} to suggest a multi-modal spatial allocation policy, where using MFD and 3D-MFD theory, the authors assessed the benefits of allowing pool ride-hailing vehicles in the bus network. 

While this macroscopic approach to model the delays of different modes in a shared network is novel to our knowledge, different delay functions for different actors and/or objectives of the actors at the link level in transportation networks have been previously analyzed. The existence of a Wardrop equilibrium for multi-class transportation network was shown in~\cite{braess1979} and~\cite{florian1977multimodal}. These works present an analysis of a multi-modal setting where public transit vehicles interact with private vehicles on the same roads. In~\cite{mehr2020autonomousrouting,lazar2021routing}, a situation where both autonomous and regular vehicles share the same route is studied, and the vehicle classes are assumed to have a different impact on the total and common delay for each link. Another type of routing game is one in which different classes of users may have different objectives, but their decisions cause delays to other classes through the use of common resources. In~\cite{nilsson2018twotier}, the authors examine a situation in which a company tries to minimize the total travel time for its fleet while interacting with regular drivers who are trying to minimize their own travel time. 

The contribution of this paper is twofold. First, we develop delay functions for multi-modal transportation networks, and use them to estimate aggregate modal- and network-dependent travel times. We also illustrate how these delay functions replicate the uncongested behavior of the MFD function. Second, we analyze the system optimum for the allocation strategy in which private vehicles and solo ride-hailing users utilize the vehicle network, while buses and ride-hailing vehicles with high occupancy use the bus network. We do so mainly to determine the network space configuration and multi-modal demand profiles for which such strategy is advantageous. Unlike in~\cite{FAYED202329}, we furthermore assess the user equilibrium, which we also refer to as Wardrop equilibrium, and advance a tolling scheme under settings where the user equilibrium and system optimum solutions do not coincide. This analysis is of particular interest because the user equilibrium we are analyzing only determines the number of solo and pool ride-hailing trips, while the system optimum considers the total multi-modal delays in both networks.    

The remainder of the paper is outlined as follows. In the next Section~\ref{sec:model}, we introduce the delay model for each mode of transportation and illustrate how this model relates to the theory of MFD. In the following Section, \ref{sec:so}, we analyze the properties of the system optimum and we define some conditions that guarantee the efficiency of the proposed strategy. In Section~\ref{sec:wardrop}, we characterize the Wardrop equilibrium for the efficient set of network configurations, and we identify solutions for which the user equilibrium and system optimum coincide naturally. In Section~\ref{sec:tolls}, we employ these results to set forward a tolling scheme for the cases where the allocation strategy is beneficial yet the Wardrop solution does not coincide with the system optimum. Numerical examples are presented in Section~\ref{sec:numerical}, and the paper concludes with some suggestions for future research.

\section{Model} \label{sec:model}
In this section, we first introduce the delay model for a multi-modal transportation network and then provide a link between this model and macroscopic traffic flow theory.

\subsection{Macroscopic Multi-modal Delay Model}
\begin{figure}
    \centering
    \begin{tikzpicture}[node distance=1cm, scale=0.27]

 \node[draw, rotate=90, minimum height=0.5cm, minimum width=4cm] (demand) at (-11,0) {Multi-modal demand};
 \node[draw, rotate=0, minimum height=1.8cm, label = Vehicle network $\mathcal{V}$, minimum width=2cm] (vehicle network) at (2.5,3.9) {};
\node[draw, rotate=0, minimum height=1.8cm, label = below:Bus network $\mathcal{B}$, minimum width=2cm] (bus network) at (2.5,-3.9) {};
 
 \node[inner sep=0pt] (bus) at (-6.5,-5)
    {\includegraphics[width=.08\textwidth]{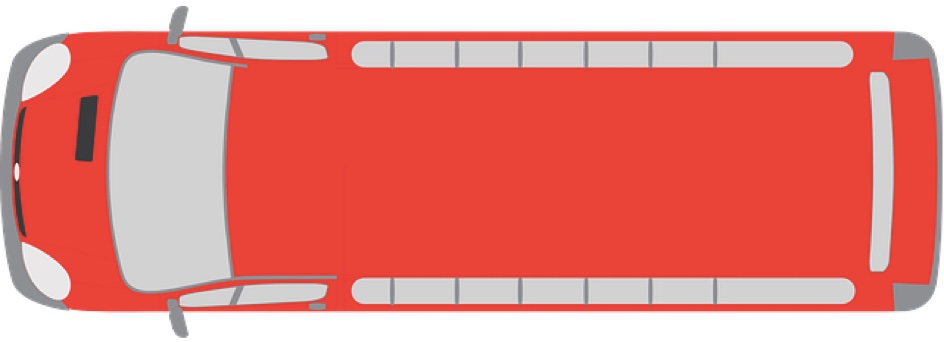}};
\node[inner sep=0pt] (car) at (-7,5)
    {\includegraphics[width=.05\textwidth]{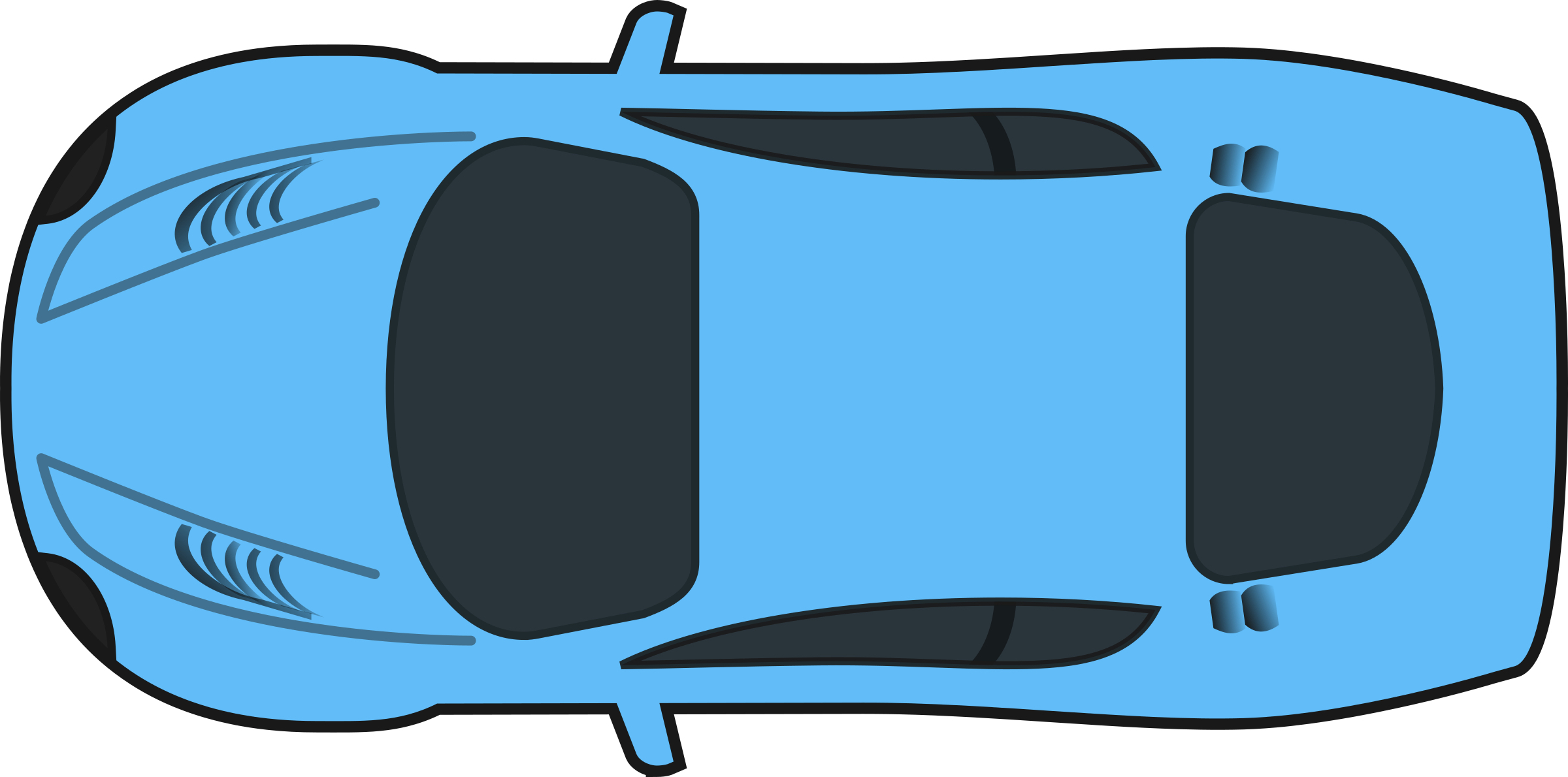}};
\node[inner sep=0pt] (rh) at (-7,0)
    {\includegraphics[width=.06\textwidth]{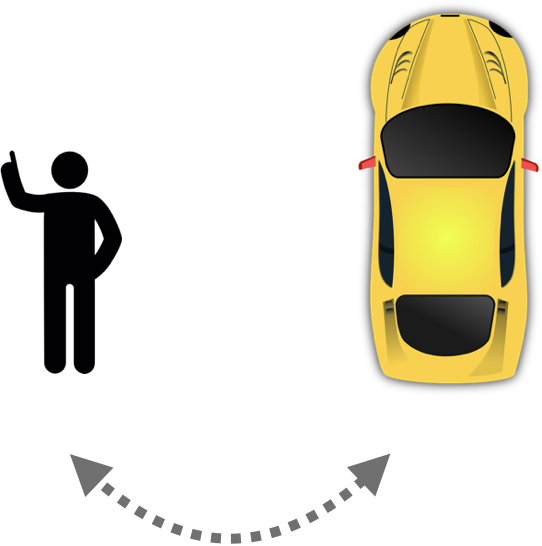}};
        \node (xpv) at (-4,5)  {};
        \node (xrs) at (-4,0)  {$\beta$};
        \node[draw, circle] (destination) at (8,0)  {};
        \node[draw, circle] (origin) at (-17,0)  {};
        \node (xb) at (-4,-5)  {};
        \draw[->] (origin) -- node[above, align = center] {$x^{rs}$} (demand);
        \draw[->] (origin) -- node[above, align = center] {$x^{pv}$} (-12,5);
        \draw[->] (origin) -- node[below, align = center] {$x^{b}$} (-12,-5);
        \draw[->] (-5,2) -- node[above, align = center] {$n_e$} (vehicle network.west);
        \draw[->] (xrs) to[bend left] node[above] {{$t_{\mathcal{V}}$}} (destination);
         \draw[<->] (vehicle network) -- node[right, align = center] {$\alpha$} (bus network);
        \draw[->,
line join=round,
decorate, decoration={
    zigzag,
    segment length=4,
    amplitude=.9,post=lineto,
    post length=2pt
}] (xrs) to[bend right] node[below] {$t_{\mathcal{B}}$} (destination);
        \draw[->] (xpv) to[bend left] node[above] {{$t_{\mathcal{V}}$}} (destination);
        \draw[->, ,
line join=round,
decorate, decoration={
    zigzag,
    segment length=4,
    amplitude=.9,post=lineto,
    post length=2pt
}] (xb) to[bend right] node[below] {$t_b$} (destination);
    \end{tikzpicture}
        \caption{Schematic sketch of the problem structure. Users travel by private vehicle, ride-hailing, or buses whose demand rates we denote as $x^{pv}$, $x^{rs}$, and $x^b$ respectively. The network space is partitioned into two parts, with a portion $\alpha \in (0,1)$ assigned to the vehicle network $\mathcal{V}$, and the remaining portion assigned to the bus network $\mathcal{B}$. Private vehicles and solo trip users utilize the vehicle network $\mathcal{V}$ with an average trip time of $t_{\mathcal{V}}$ whereas buses and pool ride-hailing vehicles travel exclusively in the bus network $\mathcal{B}$ with a travel time of $t_{\mathcal{B}}$ and $t_b$ respectively. The demand rates are exogenous but the split between solo and pool ride-hailing trips $\beta$ is endogenous.}
    \label{fig:sketch}
\end{figure}
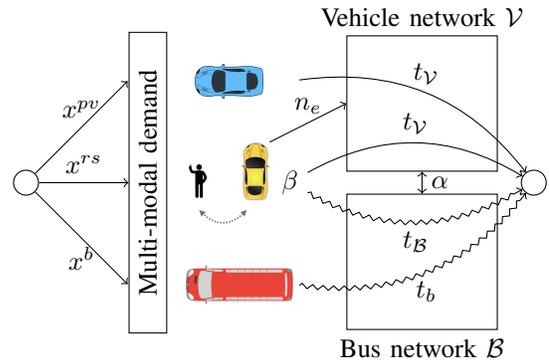

In this framework, we study a multi-modal network with private vehicles, ride-hailing, and buses, which demand we denote by $x^{pv} > 0$, $x^{rs} > 0$, and $x^b> 0$, respectively. In ride-hailing, users have the option to either ride alone or to pool, and we denote the demand for solo and pool trips by $x^{s}$ and $x^{p}$, respectively. Naturally, $x^{rs} = x^{s} + x^{p}$. The entire network infrastructure is split into a vehicle network which we denote by $\mathcal{V}$, and a bus network which we denote by $\mathcal{B}$, according to a spatial division factor $\alpha \in (0,1)$ where $\alpha$ is the space allocated for the vehicle network and $\bar{\alpha} = 1-\alpha$ is the space allocated to the bus network. The private vehicles and solo ride-hailing users are required to utilize the vehicle network, and the bus and pool ride-hailing users utilize the bus network.
Note that in this framework, we focus on pool rides with no more than two passengers. A schematic sketch of the problem setting is displayed in Figure~\ref{fig:sketch}. To further describe the proposed allocation scheme, we first define the baseline delay function that we adopt in this work. 

Let $x \in \R_{\geq 0}$ be the network flow expressed in vehicles per hour, then the average travel time in the network $t:\R_{\geq 0} \rightarrow \R_{>0}$, also called the delay function, is given by 
\begin{equation}
t(x) = t_f\biggl(1+a\left(\frac{x}{C}\right)^b\biggr) \, ,
\label{eqn:delay}
\end{equation}
where $t_f > 0$ is the free flow travel time, $C > 0$ is the network capacity, and $a > 0$ and $b > 0$ are the network-specific delay function parameters. 
The travel time in the vehicle network $\mathcal{V}$, which occupies a fraction $\alpha$ of the total network space, $t_{\mc V}: \R_{\geq 0} \times \R_{\geq 0} \times (0,1) \times \R_{\geq 0}  \rightarrow \R_{> 0}$ is given by
\begin{equation}
    t_{\mathcal{V}}(x^s, x^{pv}, \alpha, n_e) = t_f\left(1+a\left(\frac{ x^{pv} + x^{s}}{\omega(n_e) \alpha C}\right)^b\right) \, ,
    \label{eqn:vehicle_delay}
\end{equation}
where $\omega : \R_{\geq 0} \rightarrow (0,1]$ is a continuously differentiable function that depends on the number of empty ride-hailing vehicles $n_e \geq 0$. This category of vehicles is usually roaming around in the network, waiting for a pick-up request. Assuming that empty vehicles are only allowed to travel in the vehicle network $\mathcal{V}$, the purpose of $\omega$ is to capture the capacity drop in the vehicle network $\mathcal{V}$ due to the existence of idling vehicles $n_e$ in $\mathcal{V}$ such that $\frac{\d \omega}{\d n_e}\leq 0$ and $\omega(0) = 1$. The traffic flow in this network consists of the flow for private vehicles $x^{pv}$, and the flow for solo ride-hailing trips $x^s$. 

Similarly, knowing that the bus network occupies a space $\bar{\alpha} = 1-\alpha$ of the total network infrastructure, we compute the delay function for pool vehicles utilizing the bus network, $t_{\mathcal{B}}: \R_{\geq 0} \times \R_{\geq 0} \times (0,1) \rightarrow \R_{>0} $, using
\begin{equation}
    t_{\mathcal{B}}(x^p, f_b, \alpha) = t_f\left(1+a\left(\frac{\frac{x^{p}}{o^p} + f_b}{(1-\alpha) C}\right)^b\right) \Delta_p k\left(f_b\right) \, ,
    \label{eqn:pool_delay}
\end{equation}
 where the constant $o^p > 1$ is the pool vehicle occupancy, $\Delta_p > 1$ is the normalized detour factor of passengers which reflects the extra distance travelled by passengers due to them sharing their rides with other passengers. The variable $f_b > 0$ is the average bus flow in the bus network, and $k:\R_{\geq 0} \rightarrow \R_{>0}$ is a continuously differentiable function that estimates the average influence of bus flow on travel time in the bus network such that $\frac{\d k}{\d f_b}>0$ and $k(0)>1$. Within our framework, the bus flow is assumed to be constant such that buses maintain the same frequency at stops. Moreover, we assume that bus flow is always less than the bus demand such that $f_b<x^b$. The flow in the bus network consists of the pool vehicles trips equal to $\frac{x^p}{o^p}$ and the average bus flow $f_b$. 
 
 The bus user delays, $t_b :\R_{\geq 0} \times \R_{\geq 0} \times (0,1) \rightarrow \R_{>0}$ are computed in a similar manner as in \eqref{eqn:pool_delay} as shown in
\begin{equation}
t_{b}(x^p, f_b, \alpha) = t_f\left(1+a\left(\frac{\frac{x^p}{o^p} + f_b}{(1-\alpha) C}\right)^b\right) \Delta_b k\left(f_b\right) + \gamma \, ,
\label{eqn:bus_delay}
\end{equation}
except that the normalized detour ratio $\Delta_p$ is replaced by $\Delta_b$ such that $\Delta_b>1$ to account for the extra time bus users must travel compared to the direct trip, and $\gamma > 0$ is a constant describing the extra time passengers incur due to the boarding and alighting of other bus users.

In this paper, we analyze the optimal split between solo and pool rides for a given ride-hailing demand. To do so, we introduce the split $\beta \in [0,1]$ that determines the fraction of pooled rides, i.e., $x^{p} = \beta x^{rs}$ and $x^{s} = (1-\beta)x^{rs}$. The equilibrium of the system is achieved when $\beta$ is determined in two different ways: first, through the case where a central authority dictates this split, i.e., the system optimum, or second through the case where ride-hailing users attempt to minimize their own travel time when choosing between a solo trip or a pooled trip, which is commonly referred to as the Wardrop or user equilibrium. If the two equilibria are different, the Wardrop equilibrium may perform worse overall than the system optimum, and incur a so-called Price of Anarchy (PoA). In Section~\ref{sec:wardrop}, we categorize when this is not the case. Irrespectively, we note that in the particular case where the system optimum occurs for $\beta = 0$, the network configuration and the demand profile do not allow for the utilization of bus lanes by pool ride-hailing vehicles.

Since the main focus of this paper is on the influence of solo and pool rides, we will omit the dependence on these demands, i.e., with a slight abuse of notation, we let $t_{\mc V}(x^s) = t_{\mathcal{V}}(x^s, x^{pv}, \alpha, n_e),  t_{\mc B}(x^p) = t_{\mathcal{B}}(x^p, f_b, \alpha)$ and $t_{b}(x^p) = t_{b}(x^p, f_b, \alpha)$.

Before proceeding to the equilibrium analysis, we will show how the proposed delay functions relate to macroscopic traffic flow theory for urban transportation networks. 

\subsection{Relationship to Macroscopic Traffic Flow Models}

The proposed aggregate delay functions reflect modal-specific travel times at the network level. 
Under our static network equilibrium framework, these functions relate to the well-established theory of Macroscopic Fundamental Diagrams, which is often used to model urban traffic at the macroscopic level. The MFD provides a relationship between average accumulation, i.e, the total number of network vehicles, and average traffic flow in the network. It captures the relationship of the traffic dynamics on the aggregate level, and under specific settings, it can very accurately represent the network conditions. The MFD usually assumes that flow increases up to a certain level of accumulation, often referred to as the critical accumulation. Another common assumption is that the MFD is concave. The following theorem shows that the functional form of the delay function provided in~\eqref{eqn:delay} is able to capture the MFD relationship.

\begin{theorem}
When the average travel time in the network is given by~\eqref{eqn:delay}, the flow-accumulation relation $x(n)$ is increasing with accumulation $n$ and is concave with respect to this accumulation.
\end{theorem}

\medskip

\begin{proof}
    By observing that the accumulation in the network for a given flow is given by $n(x) = xt(x)$, it follows that the accumulation is strictly increasing with the flow, since
    \begin{equation}
        \frac{\d n}{\d x} = x \frac{\d t}{\d x} + t(x) > 0 \,.
    \end{equation}
    Thus, the function $n(x)$ is invertible, and let $x(n)$ be its inverse. It then follows from the inverse function rule that $x(n)$ is strictly increasing in $n$. 
    For the second part, we observe that $n(x)$ is convex with respect to $x$, since
    \begin{equation}
        \frac{\d^2 n}{\d x^2} = x\frac{\d^2 t}{\d x^2} + 2\frac{\d t}{\d x} = t_f \frac{ab(b+1)}{C}\left(\frac{x}{C}\right)^{b-1} > 0 \,.
    \end{equation}
    Then it follows from~\cite[12.21]{binmore1982mathematical} that $x(n)$ is concave. 
\end{proof}

\medskip

\begin{remark}
The MFD flow often decreases beyond a critical accumulation or density. However, in this paper we focus on the increasing regime, i.e., we assume that the average demand flow never exceeds the network capacity. 
\end{remark}

\medskip

The following example illustrates how the delay functions correspond to an MFD for one set of parameters.

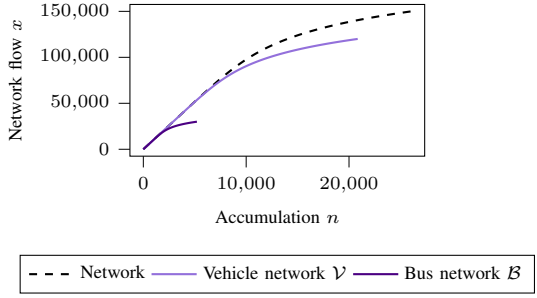
\begin{figure}
    \centering
    \input{mfds.tikz}
    \caption{The Macroscopic Fundamental Diagrams (MFDs) in Example~\ref{ex:MFD}.}
    \label{fig:MFD}
\end{figure}

\begin{example}\label{ex:MFD}
In~\eqref{eqn:delay} let $a=0.8$, $b = 6$, $C = 150000$ pax/hr, and $t_f = 0.1$ hr.
We can express the flow as a function of accumulation for the entire network under consideration, but also for the two subnetworks, by setting the network fractional split $\alpha$ to $0.8$, and multiplying the capacity by $\alpha$ and $\bar{\alpha}$ for the vehicle and bus networks, respectively. 
Figure~\ref{fig:MFD} shows that the functions obtained reproduce what we observe in large-scale networks when traffic measures are aggregated. 
This suggests that the delay function we propose in~\eqref{eqn:delay} can potentially be used to capture the relationship between flow and accumulation at the aggregate traffic level.
\end{example}
\medskip

Having established the macroscopic modelling framework for the proposed space allocation strategy, we proceed next with analyzing the system optimum properties of the multi-modal user delays.

\section{Efficiency of the System Optimum}\label{sec:so}
In the following section, we analyze the properties of the system optimum. The ultimate goal of this analysis is to determine the set $\{\alpha, x^{pv}, x^{rs}, f_b\}$ for which the system optimum yields a value of $\beta=0$, i.e., for when our proposed strategy is not able to return lower delays compared to the scenario where ride-hailing vehicles are not allowed in bus lanes. This total multi-modal delays in the network under consideration, also referred to as Passenger Hour Travelled (PHT) for all users of the network, is computed by taking into account the demand and delays of each network user. Therefore, for a given split $\beta$ and a fixed $\alpha$, the PHT is given by
 \begin{align}
    \text{PHT}(\beta) = & x^{pv}t_{\mathcal{V}}((1-\beta)x^{rs}) + (1-\beta) x^{rs}t_{\mathcal{V}}((1-\beta)x^{rs})  \\
    & + \beta x^{rs} t_{\mathcal{B}}(\beta x^{rs})  + x^bt_b(\beta x^{rs}) \, .
 \end{align}
 
The aim of the system optimum is to find the split between solo and pool demand $\beta^{\text{SO}}$ such that the total PHT of multi-modal commuters including ride-hailing users is minimized as follows
\begin{equation} 
\label{eqn:SO}
 \beta^\text{SO} \in \argmin_{\beta\in[0, 1]} \text{PHT}(\beta) \, .
\end{equation} 
The following proposition guarantees that the optimization problem~\eqref{eqn:SO} has a unique solution. 
\begin{proposition}
\label{prop:SO_unique}
If $b > 1$, then the solution to the system optimum $\beta^\text{SO}$ in~\eqref{eqn:SO} is unique.
\end{proposition}

\iftoggle{full_version}{
The proof is given in Appendix~\ref{app:proof_SO_unique}.}{The proof is given in the extended version of this paper.}
\medskip

Knowing that the minimum is unique, we can characterize how it changes with $\alpha$ for fixed private vehicles, ride-hailing, and bus demands.
\begin{proposition}\label{prop:SO_decreasing}
    If $b>1$, the solution to the system optimum $\beta^{\text{SO}}$ is a decreasing function of $\alpha$. 
\end{proposition}

\iftoggle{full_version}{
The proof is given in Appendix~\ref{app:proof_SO_decreasing}.}{The proof is given in the extended version of this paper.}
\medskip

Next, we utilize the convexity of the $\text{PHT}(\beta)$ function to identify some sufficient conditions for the system optimum

\begin{proposition}\label{prop:SO_conditions}
Suppose that $b > 1$. For a fixed demand and a given $\alpha$, the following two sufficient conditions hold:
\begin{enumerate}
        \item If $\omega \frac{\alpha}{1-\alpha}\left(\frac{x^{rs}}{o^p} + f_b\right)> \left(\frac{b+1}{\frac{b}{o^p}+1}\right)^{\frac{1}{b}}x^{pv}$ then $\beta^\text{SO} < 1$.
        \item If $\omega \frac{\alpha}{1-\alpha}f_b \geq \left(\frac{b+1}{\frac{b}{o^p}+1}\right)^{\frac{1}{b}} (x^{pv}+ x^{rs})$ then $\beta^\text{SO} = 0$.
    \end{enumerate}
Moreover, there exists an $\ul{\alpha} > 0$ such that $\beta^\text{SO} = 1$ for all $\alpha < \ul{\alpha}$.

\end{proposition}

\iftoggle{full_version}{
The proof is given in Appendix~\ref{app:proof_SO_conditions}.}{The proof is given in the extended version of this paper.}
\medskip

It can be concluded that for any space and demand configuration $\Omega$ such that $\Omega = \{ \alpha, x^{pv}, x^{rs}, f_b \mid \frac{\alpha}{1-\alpha}f_b \geq \left(\frac{b+1}{\frac{b}{o^p}+1}\right)^{\frac{1}{b}} (x^{pv}+ x^{rs})\}$, pool ride-hailing users should not be allowed in bus lanes under the system optimum solution. Nevertheless, in situations where the benefits of the pooling in bus lanes scenario is evident, i.e., when $\beta^{SO} = (0, 1]$, it is useful to look at the user's choice under this strategy. 

\section{Efficiency of the user equilibrium}\label{sec:wardrop}
We proceed in this section with the analysis of the properties of the user equilibrium, and we concertize some conditions for which this Wardrop equilibrium solution is efficient under the proposed allocation strategy by utilizing the Price of Anarchy formulation.

\subsection{Properties of the User Equilibrium}
Since the demands $x^{pv}$ and $f_b$ are strictly positive, and $n_e$ and $\alpha$ are fixed. Then, it holds that
\begin{align}
\frac{\partial t_{\mathcal{V}}}{\partial x^s} &= \frac{t_f ab}{\omega\alpha C}\left( \frac{x^{pv}+x^s}{\omega \alpha C} \right)^{b-1} > 0\,, \\
\frac{\partial t_{\mathcal{B}}}{\partial x^p} &= \frac{t_f ab}{o^p\bar{\alpha}C} \left(  \frac{\frac{x^p}{o^p} + f_b}{\bar{\alpha} C}  \right)^{b-1}\Delta_p k > 0 \, .
\end{align}
It follows from~\cite[Theorem 2.5]{patriksson2015traffic} that the Wardrop equilibrium $\beta^{\text{UE}}$ is unique and can be determined by
\begin{align}
  \beta^{\text{UE}} \in \argmin_{\beta\in[0, 1]}\int_0^{(1-\beta)x^{rs}} t_{\mathcal{V}}(s) \d s + \int_0^{\beta x^{rs}} t_{\mathcal{B}}(s) \d s \,.
  \label{eqn:UE}
\end{align}
Moreover, since the immediate consequence is that the delay functions are strictly increasing, it holds that for any $\beta\in (0, 1)$, it is a Wardrop equilibrium if and only if  $t_\mc{V}((1-\beta)x^{rs}) = t_\mc{B}(\beta x^{rs})$.

For the remainder of this section, we will investigate how the Wardrop equilibrium depends on the space allocation $\alpha$. We start with the case where ride-hailing users utilize both networks.
\begin{proposition}\label{prop:uedecreasing}
The solution to the system optimum $\beta^{\text{UE}}$ is a decreasing function of $\alpha$. 
\end{proposition}

\iftoggle{full_version}{
The proof is given in Appendix~\ref{app:proof_UE_decreasing}.}{The proof is given in the extended version of this paper.}

\medskip

While the previous proposition makes sure that a user equilibrium always decreases, the following proposition identifies some sufficient conditions for the user equilibrium.

\begin{proposition}\label{prop:sufficient_condition_wardrop}
For a fixed demand and a given $\alpha$, the following two sufficient conditions hold:
\begin{enumerate}
    \item If $(1-\alpha)x^{pv} < \omega \alpha\left(\frac{x^{rs}}{o^p} + f_b\right)$ then $\beta^\text{UE} < 1$.
    \item If $(1-\alpha)(x^{pv}+x^{rs})<\omega \alpha f_b$ then $\beta^\text{UE} = 0$.
\end{enumerate}
Moreover, there exists an $\ul{\alpha} > 0$ such that $\beta^\text{UE} = 1$ for all $\alpha < \ul{\alpha}$.
\end{proposition}

\iftoggle{full_version}{
The proof is given in Appendix~\ref{app:sufficient_conditions_UE}.}{The proof is given in the extended version of this paper.}

\medskip

\begin{remark}
Even if the average pool detour times captured by $\Delta_p$ and the slowing down of pool vehicles by buses captured by $k$ are not fixed but rather demand-dependent, the conditions in Proposition~\ref{prop:sufficient_condition_wardrop} are still valid.
\end{remark}

\subsection{Price of Anarchy}
To quantify the potential performance decrease a user optimal split inflicts on the system, we examine the Price of Anarchy (PoA), i.e., the ratio of Passengers Hours Travelled at user equilibrium to system optimum such that
\begin{equation}
    \text{PoA}(\beta^\text{UE}, \beta^\text{SO}) =  \frac{\text{PHT}(\beta^\text{UE})}{\text{PHT}(\beta^\text{SO})} \,.
\end{equation}

The following theorem summarizes some of the main findings and shows that, for certain values of $\alpha$, the need for intervention is limited since the system optimum coincides with the choice of users.

\begin{theorem}
\label{th:PoA1}
Suppose $b>1$, then there exist $\ul{\alpha} > 0$ and $\bar{\alpha} < 1$, so that for all $\alpha < \ul{\alpha}$ and all $\alpha > \bar{\alpha}$ the Price of Anarchy is $1$. Moreover, if $\beta^\text{SO} \in (0,1)$ and satisfies 
\begin{equation}
    (x^{pv}+(1-\beta^\text{SO})x^{rs})\frac{\partial t_{\mathcal{V}}}{\partial \beta} + \beta x^{rs}\frac{\partial t_{\mathcal{B}}}{\partial \beta} + x^b\frac{\partial t_b}{\partial \beta}=0\,,
\end{equation}
then the Price of Anarchy is also $1$.
\end{theorem}

\medskip

\begin{proof}
The statement about $\ul{\alpha}$ is a direct consequence of the first statements in Proposition~\ref{prop:sufficient_condition_wardrop} and Proposition~\ref{prop:SO_conditions} where $\beta^\text{UE} = \beta^\text{SO} =1$. This yields a $\text{PoA} = 1$. 

For the statement about $\bar{\alpha}$, we observe that the  condition for $\beta^\text{UE} = 0$ in Proposition~\ref{prop:sufficient_condition_wardrop} reads 
$\frac{\alpha}{1-\alpha} > \frac{x^{pv}+x^{rs}}{\omega f_b}$, 
and the condition for $\beta^\text{SO} =0$ in Proposition~\ref{prop:SO_conditions} reads 
\begin{equation}
     \frac{\alpha}{1-\alpha} \geq \frac{1}{\omega f_b}\left(\frac{b+1}{\frac{b}{o^p}+1}\right)^{\frac{1}{b}} (x^{pv}+ x^{rs})\,.
\end{equation}
Hence, since $\frac{\alpha}{1-\alpha} \rightarrow +\infty$ when $\alpha \rightarrow 1^{-}$, it is possible to find a value $\bar{\alpha} < 1$ such that $\beta^\text{UE} = \beta^\text{SO} =0$ for all $\alpha > \bar{\alpha}$.
\end{proof}

In cases of efficient allocation strategies yet inefficient user equilibrium, i.e., when $\beta^{\text{SO}} \in (0, 1]$ and $\text{PoA}>1$, tolling is required to bridge the gap between the system optimum and the user equilibrium. 

\section{Tolling} \label{sec:tolls}

In the following section, we provide a tolling scheme for when the multi-modal space allocation policy is efficient, i.e., $\beta^{\text{SO}}\in(0, 1]$, and the user equilibrium solution does not leverage the full potential of this policy, i.e., $\beta^{\text{SO}}\neq\beta^{\text{UE}}$. More specifically, we provide an additive toll to the pool user delay function $t_{\mc{B}}$ that incentivizes or deters ride-hailing users from pooling in the bus network $\mc{B}$. 

\begin{proposition}
\label{prop:toll}
    If $\tau_p \in \R$ is the toll for utilizing the bus lanes by pool ride-hailing users, then by letting 
    \begin{multline}\label{eqn:toll}
        \tau_p =     -t_fab\left( \frac{x^{pv}+(1-\beta^\text{SO})x^{rs}}{\omega \alpha C}  \right)^b \\ + \frac{ t_f ab k}{o^p\bar{\alpha}C}\left( \frac{\beta^{\text{SO}} \frac{x^{rs}}{o^p} + f_b}{\bar{\alpha} C} \right)^{b-1}(x^{rs}\Delta_p + x^b \Delta_b)\, ,
    \end{multline}
        the socially optimal solution is achieved. 
\end{proposition}

\iftoggle{full_version}{
The proof is given in Appendix~\ref{app:toll}.}{The proof is given in the extended version of this paper.}

\medskip
\begin{remark}
    We note that the expression of tolling holds for $\beta^{\text{SO}} = 1$ and is intrinsically equal to $0$ when $\beta^{\text{SO}} = \beta^{\text{UE}}$ and $\beta^{\text{SO}}\in (0,1)$. The latter point is straightforwardly observed from Theorem~\ref{th:PoA1} which leads to $\tau_p = 0$. The former is the outcome of looking at the necessary conditions for when $\beta^{\text{SO}} = 1$. Therefore, when all ride-hailing users opt for pooling, we have that 

    \begin{equation}
    \label{eqn:toll_analysis}
        x^{pv}\frac{\partial t_{\mc{V}}}{\partial \beta} + x^{rs}\frac{\partial t_{\mc{B}}}{\partial \beta} + x^b\frac{\partial t_{b}}{\partial \beta}\leq x^{rs}(t_{\mc{V}} - t_{\mc{B}}) \, .
    \end{equation}

When $t_{\mc{V}}< t_{\mc{B}}$ for $\beta^{\text{SO}}=1$, then we have that $\beta^{\text{UE}} = 0$ at system optimum. Introducing a toll $\tau_p$ is therefore necessary to yield $x^{rs}\tau_p - x^{rs}(t_{\mc{V}} - t_{\mc{B}}) \leq 0$, which implies that $\beta^{\text{UE}} = 1$. Nevertheless, when $t_{\mc{V}}\geq t_{\mc{B}}$, tolls are not needed. However, adding $\tau_p$ to the cost of pool ride-hailing users will not modify their choices. Therefore, the tolling function is still valid though unnecessary.

\end{remark}

\section{Numerical Examples}\label{sec:numerical}

Next, we illustrate the efficiency of the space allocation strategy with a numerical example and compare the optimal solo and pool ride-hailing split for the system optimum and the user equilibrium for different values of the network split $\alpha$. 

For this purpose, we use the delay function defined in Example~\ref{ex:MFD}. In terms of the constants, we set the pool and bus trip detour factors $\Delta_p$ to $1.2$ and $\Delta_b$ to $1.4$ respectively, the additional travel time of bus users due to stopping at stations $\gamma$ to $0.05$ hr, the average pool trip occupancy $o^p$ to $1.6$ pax/veh, and the bus flow $f_b$ to $12000$ bus/hr. Note that despite considering pool trips with no more than two passengers, the pool vehicle occupancy is less than $2$ because users do not have exactly the same origins and destinations. Furthermore, we assume that the number of idle vehicles $n_e$ is constant, so $\omega(n_e) = 0.97$ captures the capacity drop in network $\mc{V}$ due to $n_e$. To account for the potential of buses to slow down pool vehicles due to their frequent stops, we set the factor $k$ to $1.15$.
The example under consideration assumes that the private vehicle $x^{pv}$, bus $x^b$, and ride-hailing $x^{rs}$ demand are equal to $80000$, $35000$, and $100000$ pax/hr respectively. We note that the result section only shows solutions for values of $\alpha$ where the capacity limits are not exceeded in both the vehicle and bus networks.

\begin{figure}
    \centering
    \input{SO_UE_Ben.tikz}\input{betas.tikz}
    \caption{Comparison of PHT and $\beta$ for $x^{pv} = 80000$, $x^{rs} = 35000$, and $x^b = 100000$.}
    \label{fig:SO_UE_Ben}
\end{figure}
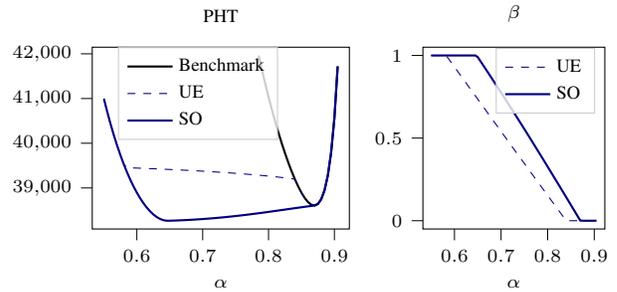

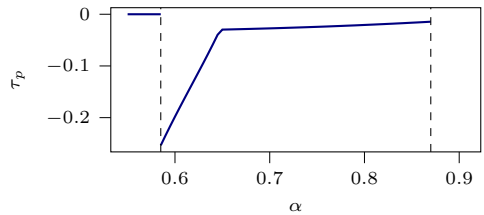
\begin{figure}
    \centering
    \input{fares.tikz}
    \caption{Tolls for the different values of $\alpha$}
    \label{fig:tolls}
\end{figure}

Figure~\ref{fig:SO_UE_Ben} displays the results of the PHT under the benchmark (BM), the user equilibrium, and the system optimum scenarios for different network configurations $\alpha$. Note that the benchmark scenario describes ordinary network settings where all ride-hailing users utilize the vehicle network~$\mc{V}$, i.e., $\beta = 0$. It can be observed that for high values of $\alpha$, the three different curves merge together, reflecting that when the network space allocated to buses is relatively compact, pooling must not be allowed in bus lanes to avoid penalizing bus users. This coincides with ride-hailing users' choice who find utilizing the bus network very costly compared to travelling solo in the vehicle network. On the contrary, for relatively low values of $\alpha$, the PHT curves merge because the large network space allocated for buses implies that pooling in bus lanes is a convenient solution from both a system optimum and user equilibrium perspective. For intermediate values of $\alpha$, we observe a gap between the PHT curves. This gap can be explained by looking at the values of $\beta$ in Figure~\ref{fig:SO_UE_Ben}. Since a pool trip length is distance-wise longer, ride-hailing users prefer the solo over the pool option, which substantiates why $\beta^{\text{UE}}\leq\beta^{\text{SO}}$ for the different space configuration values $\alpha$. Therefore, a tolling scheme is needed to mostly incentivize ride-hailing users to pool in bus lanes, i.e., the tolls are in fact discounts. The toll values are displayed in Figure~\ref{fig:tolls} where it can be seen that for low values of $\alpha$, no tolling is needed as the system optimum and user equilibrium solutions occur naturally for $\beta = 1$, while for high values of $\alpha$, the space allocation strategy proposed is inefficient and should be therefore not activated. 

To further understand the efficiency of our strategy, we compare in Table~\ref{tab:PHT_comparison} the PHT values for private vehicles $\text{PHT}\, pv$, ride-hailing users $\text{PHT} \, rs$, and $\text{PHT} \, b$ for $\alpha = 0.869$ and $\alpha = 0.647$. The choice of these values corresponds to the network configuration where the delays are minimal under the benchmark and the system optimum scenario. While the total demand under the benchmark scenario exceeds the network capacity for $\alpha=0.647$, we observe an improvement in total delays under the SO settings compared to scenarios where $\alpha = 0.869$ where the benchmark, the user equilibrium, and the system optimum mostly yield a solution with no pooling in bus lanes.

\begin{table}
\caption{PHT for different $\alpha$}
\begin{tabular}{lccc|ccc}
&\multicolumn{3}{c}{$\alpha = 0.869$} & \multicolumn{3}{c}{$\alpha = 0.647$} \\
\hline
 & BM & UE & SO &BM & UE& SO \\
\hline
    PHT ${pv}$ &$11053$ & $11053$& $11018$& -& $11884$& $9937$  \\
    PHT ${rs}$ &$4836$ & $4836$&  $4823$&- & $5199$& $5383$  \\
    PHT ${b}$ &$22717$ & $22717$&$22763$ & -&$22331$ & $22944$ \\ 
\hline
Total & $38606$ & $38606$ & $38604$ & - & $39414$ & $38264$\\ \hline
\label{tab:PHT_comparison}
\end{tabular}
\end{table}

\section{Conclusions}
In this work, we propose a space allocation policy where pool ride-hailing users are allowed to travel in the bus network to compensate for the extra detour caused by pooling, while solo users perform their trips in the vehicle network concurrently with private vehicles. We use macroscopic-level delay functions to estimate aggregate modal and network-dependent travel times. We also show that this approach replicates well the static analysis with network level MFDs.
We then assess and compare the system optimum and user equilibrium, narrowing down the user choice to solo and pool for the user equilibrium, while considering the compounded delays for all network users for the system optimum. Finally, we propose a pricing scheme to ensure the efficiency of the user equilibrium in instances where the space allocation scheme proposed reduces overall multi-modal user delays.  

In the future, we plan to investigate a similar space allocation policy for pool trips with more than two passengers, where the pool vehicle occupancy and pool detour are demand-dependent.
Moreover, we aim to capture in future work how the number of idle vehicles changes with the ride-hailing demand.

\bibliographystyle{IEEEtran}
\bibliography{references.bib}

\iftoggle{full_version}{
\appendix

\subsection{Proof of Proposition~\ref{prop:SO_unique}} \label{app:proof_SO_unique}

\begin{proof}
The second derivative of PHT with respect to $\beta$ is 
\begin{multline}
\frac{\partial^2 \text{PHT}}{\partial \beta^2 } = x^{pv}\frac{\partial^2 t_{\mathcal{V}}}{\partial \beta^2} + (1-\beta)x^{rs} \frac{\partial^2 t_{\mathcal{V}}}{\partial \beta^2}-  
2x^{rs}\frac{\partial t_{\mathcal{V}}}{\partial \beta} + \\ \beta x^{rs}\frac{\partial^2 t_{\mathcal{B}}}{\partial \beta^2} + 2x^{rs} \frac{\partial t_{\mathcal{B}}}{\partial \beta} + x^b\frac{\partial^2 t_b}{\partial \beta^2}\, ,
\end{multline}
where $\frac{\partial t_{\mathcal{V}}}{\partial \beta} = -\frac{t_f abx^{rs}}{\omega \alpha C}\left(  \frac{x^{pv} + (1-\beta)x^{rs}}{\omega \alpha C}  \right)^{b-1}<0
$ and $\frac{\partial t_{\mathcal{B}}}{\partial \beta} = \frac{t_f abx^{rs}}{o^p\bar{\alpha}C} \left( \beta \frac{\frac{x^{rs}}{o^p} + f_b}{\bar{\alpha} C}  \right)^{b-1}\Delta_p k>0$. The second derivative of the delays with respect to $\beta$ for both $ t_{\mathcal{V}}$ and $t_{\mathcal{B}}$ is $\frac{\partial^2 t_{\mathcal{V}}}{\partial \beta^2} = \frac{t_f ab(b-1)x^{rs2}}{(\omega \alpha C)^2}\left(  \frac{x^{pv} + (1-\beta)x^{rs}}{\omega \alpha C}  \right)^{b-2}>0$ and $\frac{\partial^2 t_{\mathcal{B}}}{\partial \beta^2} = \frac{t_f ab(b-1)x^{rs2}}{(o^p\bar{\alpha}C)^2} \left( \beta \frac{\frac{x^{rs}}{o^p} + f_b}{\bar{\alpha} C}  \right)^{b-2}\Delta_p k>0$ respectively, since $b > 1$. The second derivative of $t_b$ with respect to $\beta$ is the same as $\frac{\partial^2 t_{\mathcal{B}}}{\partial \beta^2}$, except for $\Delta_p$ that is replaced by $\Delta_b$. We conclude that $\frac{\partial^2 \text{PHT}}{\partial \beta^2 }>0$ indicating that the PHT is strictly convex with respect to $\beta$.
\end{proof}

\subsection{Proof of Proposition~\ref{prop:SO_decreasing}}
\label{app:proof_SO_decreasing}
\begin{proof}
    Starting with the necessary condition for the system optimum when $\beta \in (0,1)$,  $\frac{\partial \text{PHT}}{\partial \beta} = 0$, we get
    \begin{multline}
    \label{eqn:so_cond}
          (x^{pv}+ (1-\beta^{\text{SO}})x^{rs})\frac{\partial t_{\mathcal{V}}}{\partial \beta} + \beta^{\text{SO}}x^{rs} \frac{\partial t_{\mathcal{B}}}{\partial \beta} +  x^b\frac{\partial t_b}{\partial \beta} \\= x^{rs}(t_{\mathcal{V}} - t_{\mathcal{B}}) \, ,
    \end{multline}
 and computing the implicit derivative $\frac{\partial \beta^\text{SO}}{\partial \alpha}$, we get that 
\begin{multline}
    \frac{\partial \beta^\text{SO}}{\partial \alpha} =  -\frac{1}{x^{rs}E} \biggl( \frac{A^b(b+1)}{\alpha} + \frac{B^b\Delta_pk}{\bar{\alpha}} +\\ \left( \frac{k B^{b-1}}{o^p\bar{\alpha}^2C} + \frac{k(b-1)B^{b-1}}{o^p\bar{\alpha}^2C}\right) (x^b\Delta_b + \beta^{\text{SO}}x^{rs}\Delta_p)\biggr)\,,
\end{multline}
 with $A = \frac{x^{pv} + (1-\beta^{\text{SO}})x^{rs}}{\omega \alpha C}$, $B=\frac{\beta^{\text{SO}} \frac{x^{rs}}{o^p} + f_b}{\bar{\alpha}C}$, and
 \begin{multline}
 E ={A^{b-1}(b+1)}{\omega\alpha C} + \frac{B^{b-1}\Delta_pk}{o^p\bar{\alpha}C}+  \\\frac{k\Delta_pB^{b-1}}{o^p\bar{\alpha}C}+ \frac{k(b-1)B^{b-2}}{(o^p\bar{\alpha}C)^2}(x^b\Delta_b + \beta^{\text{SO}}x^{rs}\Delta_p) \,.
 \end{multline}
Since $A> 0, B > 0$ and $E>0$ due to the fact that $b > 1$, it follows that $\frac{\partial \beta^\text{SO}}{\partial \alpha} < 0$. Moreover, for $\beta = 1$, we observe that $\text{PHT}(1) \rightarrow +\infty $ when $\alpha \rightarrow 0^+$, which concludes the proof.
 \end{proof}

 \medskip

 \subsection{Proof of Proposition~\ref{prop:SO_conditions}}
 \label{app:proof_SO_conditions}
\begin{proof}
We start by observing that
\begin{multline}\label{eqn:dPHT}
    \frac{\partial \text{PHT}}{\partial \beta} = -t_fabx^{rs}\left( \frac{x^{pv} + (1-\beta)x^{rs}}{\omega \alpha C}\right)^b - \\t_fx^{rs}\left( 1+ a \left( \frac{x^{pv} + (1-\beta)x^{rs}}{\omega \alpha C}\right)^b \right) + \\ t_f x^{rs}\left(1+a \left( \frac{\beta\frac{x^{rs}}{o^p} + f_b}{\bar{\alpha}C} \right)^b \right)\Delta_pk + \\ \frac{t_fabx^{rs}}{o^p\bar{\alpha}C} \left( \frac{\beta\frac{x^{rs}}{o^p} + f_b}{\bar{\alpha}C} \right)^{b-1} \left( x^b\Delta_b + \beta x^{rs}\Delta_p \right) \, .
\end{multline}

Due to the convexity of $\text{PHT}(\beta)$,  $\beta^\text{SO} = 1$ implies $\frac{\partial \text{PHT}}{\partial \beta}(1) \leq 0$. Hence, by utilizing~\eqref{eqn:dPHT}, we get that 
\begin{multline}
-ab\left(\frac{x^{pv}}{ \omega \alpha C}\right)^b + \frac{x^{rs}ab\Delta_p k}{o^p\bar{\alpha}C}\left(\frac{\frac{x^{rs}}{o^p}+f_b}{\bar{\alpha}C}\right)^{b-1} \\ 
+ \frac{x^bab\Delta_b k}{o^p\bar{\alpha}C}\left(\frac{\frac{x^{rs}}{o^p}+f_b}{\bar{\alpha}C}\right)^{b-1} -a\left(\frac{x^{pv}}{\omega \alpha C}\right)^b \\ 
+a\left(\frac{\frac{x^{rs}}{o^p} + f_b}{\bar{\alpha}C}\right)^{b}\Delta_b k \leq1-\Delta_pk \, .   
\end{multline}
Since $\Delta_pk>1$, we have that $1-\Delta_pk<0$, and therefore
\begin{multline}
\frac{b}{o^p\bar{\alpha}C}\left(\frac{\frac{x^{rs}}{o^p} + f_b}{\bar{\alpha}C}\right)^{b-1}(x^{rs}\Delta_pk + x^b\Delta_bk) \\
+ \left(\frac{\frac{x^{rs}}{o^p} + f_b}{\bar{\alpha}C}\right)^{b} 
\leq (b+1)\left(\frac{x^{pv}}{\omega \alpha C}\right)^b \, .
\end{multline}
Since $f_b<x^b$, $\Delta_pk>1$, and $\Delta_bk>1$, it follows that $\left(\frac{b}{o^p} + 1\right)\left(\frac{\frac{x^{rs}}{o^p} + f_b}{\bar{\alpha}C}\right)^{b}\leq (b+1)\left(\frac{x^{pv}}{\omega \alpha C}\right)^b$.
Finally, since $b>$, we get that $\omega \frac{\alpha}{1-\alpha}\left(\frac{x^{rs}}{o^p} + f_b\right)\leq \left(\frac{b+1}{\frac{b}{o^p}+1}\right)^{\frac{1}{b}}x^{pv}$.
Therefore, we know that if $ \omega \frac{\alpha}{1-\alpha}\left(\frac{x^{rs}}{o^p} + f_b\right)> \left(\frac{b+1}{\frac{b}{o^p}+1}\right)^{\frac{1}{b}}x^{pv}$, then the system optimum can not occur for $\beta=1$.

For the second statement, starting with $\beta^\text{SO} > 0$ implies $\frac{\partial \text{PHT}}{\partial \beta}(0)<0$, we get that
\begin{multline}
          -ab\left(\frac{x^{pv}+x^{rs}}{ \omega \alpha C}\right)^b + \frac{x^bab\Delta_b k}{o^p\bar{\alpha}C}\left(\frac{f_b}{\bar{\alpha}C}\right)^{b-1} \\ -a\left(\frac{x^{pv} + x^{rs}}{\omega \alpha C}\right)^b + a\left(\frac{f_b}{\bar{\alpha}C}\right)^b\Delta_pk <1-\Delta_pk \, .
\end{multline}
Since $1-\Delta_pk<0$, it follows that 
\begin{multline}
     \frac{x^bb\Delta_b k}{o^p\bar{\alpha}C}\left(\frac{f_b}{\bar{\alpha}C}\right)^{b-1} + \left(\frac{f_b}{\bar{\alpha}C}\right)^b\Delta_pk   < \\ b\left(\frac{x^{pv}+x^{rs}}{ \omega \alpha C}\right)^b+ \left(\frac{x^{pv} + x^{rs}}{\omega \alpha C}\right)^b  \, .
\end{multline}
Since $f_b<x^b$ and $\Delta_pk>1$, we get that
\begin{equation}
     \left(\frac{b}{o^p}+1\right)\left(\frac{f_b}{\bar{\alpha}C}\right)^b < (b+1)\left(\frac{x^{pv}+ x^{rs}}{\omega \alpha C}\right)^b \, .
\end{equation}
Finally, by utilizing that $b>1$, we obtain that 
\begin{equation}
    \omega \frac{\alpha}{1-\alpha}f_b < \left(\frac{b+1}{\frac{b}{o^p}+1}\right)^{\frac{1}{b}} (x^{pv}+ x^{rs}) \, .
\end{equation}
Therefore, we know that if $\omega \frac{\alpha}{1-\alpha}f_b \geq \left(\frac{b+1}{\frac{b}{o^p}+1}\right)^{\frac{1}{b}} (x^{pv}+ x^{rs})$, then $\beta^\text{SO} = 0$.

To prove the last part, we observe that for a fixed set of demands and all $\beta \in [0,1]$, it holds that when ${\alpha\to 0^+}$ then $\frac{\partial \text{PHT}}{\partial \beta} \to -\infty$ and hence $\beta^\text{SO} = 1$. This combined with Proposition~\ref{prop:SO_decreasing} proves the statement. 
\end{proof}

\medskip

 \subsection{Proof of Proposition~\ref{prop:uedecreasing}}

\label{app:proof_UE_decreasing}
\begin{proof}
We start with the observation that if $\beta^{\text{UE}} \in (0,1)$, then it must hold that  $t_{\mathcal{V}} = t_{\mathcal{B}}$. Therefore, 
\begin{multline}
    1+a\left(\frac{x^{pv} + (1-\beta^{\text{UE}})x^{rs}}{\omega\alpha C}\right)^b = \\
    \left(1+a\left(\frac{\beta^{\text{UE}}\frac{x^{rs}}{o^p} + f_b}{\bar{\alpha} C}\right)^b\right) \Delta_p k\left(f_b\right) \, .
\end{multline}

By taking the implicit derivative of the above expression with respect to $\alpha$, we obtain that
\begin{multline}
 \frac{\partial \beta^{\text{UE}}}{\partial \alpha} = \\ 
 - \frac{\frac{1}{\alpha}\left(\frac{x^{pv} + (1-\beta^{\text{UE}})x^{rs}}{\omega\alpha C}\right)^b + \frac{1}{\bar{\alpha}}\Delta_pk\left(\frac{\beta^{\text{UE}}\frac{x^{rs}}{o^p} + f_b}{\bar{\alpha} C}\right)^b}{\frac{x^{rs}}{\omega \alpha C}\left(\frac{x^{pv} + (1-\beta^{\text{UE}})x^{rs}}{\omega \alpha C}\right)^{b-1} + \frac{x^{rs}}{o^p\bar{\alpha}C} \Delta_p k \left(\frac{\beta^{\text{UE}}\frac{x^{rs}}{o^p} + f_b}{\bar{\alpha} C}\right)^{b-1} } \, ,   
\end{multline}
which is always negative because all the variables in the expression are strictly positive. Moreover, for $\beta = 1$, we observe that $\text{PHT}(1) \rightarrow +\infty $ when $\alpha \rightarrow 0^+$, implying that the user equilibrium solution $\beta^{\text{UE}}$ always decreases with~$\alpha$.
\end{proof}

\medskip

\subsection{Proof of Proposition~\ref{prop:sufficient_condition_wardrop}}
\label{app:sufficient_conditions_UE}
\begin{proof}
To prove the first part of the statement, we will show that $\beta^\text{UE}=1$ implies that $ (1-\alpha)x^{pv} \geq \omega \alpha\left(\frac{x^{rs}}{o^p} + f_b\right)$. In fact, if $\beta^{\text{UE}} = 1$, it must hold that the solo trip delays in the vehicle network $t_{\mathcal{V}}$ are greater than or equal to the pool trip delays in the bus network $t_{\mathcal{B}}$, and therefore
\begin{equation}
t_f\biggl(1+a\left(\frac{x^{pv}}{\omega\alpha C}\right)^b\biggr) \geq t_f\left(1+a\left(\frac{\frac{x^{rs}}{o^p} +  f_b} {\bar{\alpha} C}\right)^b\right) \Delta_p k \, .
\end{equation}
Since $t_f > 0$ and $\Delta_p k>1$, we get that
\begin{equation}
\left(\frac{x^{pv}}{\omega \alpha C}\right)^b \geq  \left(\frac{\frac{x^{rs}}{o^p} +  f_b} {\bar{\alpha} C}\right)^b \,.
\end{equation}
Moreover, since $b > 0$, it follows that
$(1-\alpha)x^{pv}\geq\omega\alpha\left(\frac{x^{rs}}{o^p} +  f_b\right)$.
Hence, $ (1-\alpha)x^{pv}<\omega\alpha\left(\frac{x^{rs}}{o^p} +  f_b\right)$ implies $\beta^\text{UE} < 1$.

For the second part of the statement, starting from  $(1-\alpha)(x^{pv}+x^{rs})<\omega \alpha f_b$, we get that
$
\frac{x^{pv}+x^{rs}}{\omega \alpha C}<\frac{f_b}{\bar{\alpha}C}$.
Since $a>0$ and $b>0$, it follows that
\begin{equation}
1+a\left(\frac{x^{pv}+x^{rs}}{\omega \alpha C}\right)^b <1+a\left(\frac{f_b}{\bar{\alpha}C}\right)^b \, .
\end{equation}
Given that $t_f>0$ and $\Delta_p k>1$, then 
\begin{equation}
t_f\left(1+a\left(\frac{x^{pv}+x^{rs}}{\omega\alpha C}\right)^b\right)<t_f\left(1+a\left(\frac{f_b}{\bar{\alpha}C}\right)^b\right)\Delta_p k \, ,
\end{equation}
which implies that $t_{\mathcal{V}} (x^{rs}) <t_{\mathcal{B}} (0)$.

To prove the last part of the statement, we first show the existence of a value of $\alpha$ such that $t_{\mc V}((1-\beta)x^{rs}) < t_{\mc B}(\beta x^{rs})$ for all $\beta \in [0,1]$. Since for a fixed set of demands and every choice of $\beta$, it holds that $t_{\mc V}((1-\beta)x^{rs}) \rightarrow +\infty$ when $\alpha \rightarrow 0^+$, such an $\alpha$ must exist. This together with Proposition~\ref{prop:uedecreasing} proves the last part.
\end{proof}

\medskip

 \subsection{Proof of Proposition~\ref{prop:toll}}

\label{app:toll}
\begin{proof}
A split between solo and pool $\beta$ is an interior solution for the system optimum if and only if it satisfies 
\begin{equation}
          (x^{pv}+ (1-\beta)x^{rs})\frac{\partial t_{\mathcal{V}}}{\partial \beta} + \beta x^{rs} \frac{\partial t_{\mathcal{B}}}{\partial \beta} +  x^b\frac{\partial t_b}{\partial \beta} = x^{rs}(t_{\mathcal{V}} - t_{\mathcal{B}}) \, .
    \end{equation}

By adding a toll to the utilization of pool users in bus lanes, a user equilibrium split between solo and pool $\beta^{\text{UE}}$ is given by
\begin{equation}
  \beta^{\text{UE}} \in \argmin_{\beta\in[0, 1]} \int_0^{(1-\beta)x^{rs}} t_{\mathcal{V}}(s) \d s + \int_0^{\beta x^{rs}} t_{\mathcal{B}}(s) + \tau_p \d s \,.   
\end{equation}
Differentiating the two integrals, and using the uniqueness property of the solution, we get that a solution to the user equilibrium problem with tolling satisfies
$x^{rs}\tau_p = x^{rs} (t_{\mc{V}} - t_{\mc{B}})$. Therefore, setting $\tau_p = (\frac{x^{pv}}{x^{rs}}+(1-\beta^\text{SO}))\frac{\partial t_{\mathcal{V}}}{\partial \beta} + \frac{\partial t_{\mathcal{B}}}{\partial \beta} + \frac{x^b}{x^{rs}}\frac{\partial t_b}{\partial \beta}$, we recover the necessary condition for the system optimum from~\eqref{eqn:so_cond}. Rewriting the toll function and replacing the derivatives with their expressions, we get~\eqref{eqn:toll}.
\end{proof}
}{}

\end{document}

%% file: mfds.tikz
\begin{tikzpicture}

\begin{axis}[
width=5.5cm,
height=3.6cm,
legend style={at={(0.5,-0.65)},anchor=north},
legend columns=3,
tick align=outside,
tick pos=left,
xlabel={\scriptsize Accumulation $n$},
xmin=-1302.71815554039, xmax=27357.0812663483,
xtick style={color=black},
ylabel={\scriptsize Network flow $x$},
ymin=-7499.95, ymax=157498.95,
ytick style={color=black},
ylabel near ticks,
scaled y ticks = false,
y tick label style={/pgf/number format/fixed, font=\scriptsize},
xlabel near ticks,
scaled x ticks = false,
x tick label style={/pgf/number format/fixed, font=\scriptsize}
]

\addplot [thick, black, dashed]
table {%
0 0
5125.8759765625 53035
6085.61279296875 62793
6785.7177734375 69754
7355.171875 75259
7851.02783203125 79898
8291.1826171875 83869
8703.1884765625 87443
9087.41015625 90640
9457.6962890625 93590
9818.9990234375 96341
10174.7685546875 98926
10521.32421875 101327
10868.1748046875 103617
11210.8388671875 105772
11551.8359375 107815
11892.5439453125 109760
12234.3955078125 111620
12583.0390625 113428
12934.4365234375 115165
13289.5439453125 116839
13644.4931640625 118436
14004.6923828125 119984
14370.64453125 121487
14742.4150390625 122947
15120.8916015625 124369
15506.2626953125 125755
15898.7470703125 127107
16298.5927734375 128427
16706.412109375 129718
17118.548828125 130970
17539.134765625 132197
17971.931640625 133410
18409.931640625 134590
18860.716796875 135758
19316.734375 136895
19778.560546875 138004
20253.951171875 139104
20735.408203125 140178
21229.908203125 141242
21731.24609375 142283
22246.513671875 143316
22768.583984375 144327
23304.53125 145330
23850.048828125 146317
24403.125 147285
24973.48828125 148251
25551.716796875 149199
26054.36328125 149999
};
\addlegendentry{ \scriptsize Network}
\addplot [thick, color1]
table {%
0 0
4234.89111328125 43802
5040.07666015625 51955
5624.33837890625 57712
6098.955078125 62233
6509.25830078125 65993
6884.27392578125 69286
7233.7939453125 72217
7567.69775390625 74884
7889.06298828125 77324
8202.625 79584
8514.6357421875 81716
8820.158203125 83694
9124.0166015625 85558
9428.19140625 87326
9733.9267578125 89010
10042.6962890625 90622
10355.2197265625 92169
10672.3427734375 93658
10994.833984375 95095
11319.7236328125 96470
11651.3134765625 97804
11990.013671875 99100
12332.91796875 100349
12680.3662109375 101555
13036.2138671875 102733
13400.8818359375 103885
13771.4521484375 105003
14150.8349609375 106097
14539.9052734375 107170
14935.8056640625 108215
15341.22265625 109240
15756.7998046875 110247
16182.841796875 111237
16616.48046875 112204
17060.556640625 113155
17509.962890625 114080
17975.779296875 115002
18447.314453125 115900
18929.9375 116785
19427.578125 117664
19938.755859375 118534
20456.751953125 119384
20843.36328125 119999
};
\addlegendentry{\scriptsize Vehicle network $\mathcal{V}$}
\addplot [thick, color2]
table {%
0 0
1296.07666015625 13348
1554.83935546875 15838
1750.59838867188 17573
1921.47387695312 18949
2081.0341796875 20107
2235.9248046875 21116
2390.04052734375 22016
2545.81713867188 22832
2705.15405273438 23582
2869.056640625 24277
3038.71240234375 24927
3214.95532226562 25539
3398.33569335938 26118
3589.65942382812 26669
3789.28344726562 27195
3997.71704101562 27699
4215.67529296875 28184
4444.12939453125 28653
4683.26708984375 29107
4932.708984375 29546
5194.47705078125 29974
5210.36328125 29999
};
\addlegendentry{\scriptsize Bus network $\mathcal{B}$}
\end{axis}

\end{tikzpicture}

%% file: SO_UE_Ben.tikz
\begin{tikzpicture}

\definecolor{color0}{rgb}{0.75,0,0.75}

\begin{axis}[
width=5cm,
height=4cm,
legend cell align={left},
legend style={
  fill opacity=0,
  draw opacity=1,
  text opacity=1,
  at={(0.1,0.49)},
  anchor=south west,
  draw=black
},
legend cell align={left},
legend style={fill opacity=0.8, draw opacity=1, text opacity=1, draw=white!80!black},
tick align=outside,
tick pos=left,
tick align=outside,
tick pos=left,
unbounded coords=jump,
x grid style={white!69.0196078431373!black},
xlabel={\scriptsize \(\displaystyle \alpha\)},
xmin=0.53225, xmax=0.92275,
xtick style={color=black},
y grid style={white!69.0196078431373!black},
title={\scriptsize PHT},
ymin=38082.0552672739, ymax=42147.9025389707,
ytick style={color=black},
ylabel near ticks,
scaled y ticks = false,
y tick label style={/pgf/number format/fixed, font = \scriptsize},
xlabel near ticks,
scaled x ticks = false,
x tick label style={/pgf/number format/fixed, font = \scriptsize},
]
\addplot [thick, black]
table {%
0.55 nan
0.555 nan
0.56 nan
0.565 nan
0.57 nan
0.575 nan
0.58 nan
0.585 nan
0.59 nan
0.595 nan
0.6 nan
0.605 nan
0.61 nan
0.615 nan
0.62 nan
0.625 nan
0.63 nan
0.635 nan
0.64 nan
0.645 nan
0.65 nan
0.655 nan
0.66 nan
0.665 nan
0.67 nan
0.675 nan
0.68 nan
0.685 nan
0.69 nan
0.695 nan
0.7 nan
0.705 nan
0.71 nan
0.715 nan
0.72 nan
0.725 nan
0.73 nan
0.735 nan
0.74 nan
0.745 nan
0.75 nan
0.755 nan
0.76 nan
0.765 nan
0.77 nan
0.775 nan
0.78 nan
0.785 41963.0912993481
0.79 41638.9642485581
0.795 41330.1153469698
0.8 41036.0529607255
0.805 40756.376158542
0.81 40490.7869621752
0.815 40239.1073922176
0.82 40001.302952914
0.825 39777.5148594898
0.83 39568.1042625724
0.835 39373.7131091194
0.84 39195.3483170907
0.845 39034.4989733901
0.85 38893.3008300722
0.855 38774.7693339119
0.86 38683.1331766262
0.865 38624.3171993069
0.87 38606.6502792055
0.875 38641.9171366321
0.88 38746.9442233095
0.885 38946.0291786895
0.89 39274.7273002784
0.895 39785.8647395735
0.9 40559.2852666378
0.905 41718.0064815411
};
\addlegendentry{\scriptsize Benchmark}
\addplot [dashed, blue!50.1960784313725!black]
table {%
0.55 40995.0465039258
0.555 40704.010125904
0.56 40433.5629362008
0.565 40182.5420405061
0.57 39949.9024182805
0.575 39734.7102652148
0.58 39536.137404263
0.585 39452.5851339595
0.59 39449.8760438966
0.595 39447.1460799554
0.6 39444.1310763153
0.605 39441.3576085315
0.61 39438.5611730851
0.615 39435.478403065
0.62 39432.3717263347
0.625 39429.5022948078
0.63 39426.3449937657
0.635 39423.1612236036
0.64 39419.9500550607
0.645 39416.4498478969
0.65 39413.1812566998
0.655 39409.6222460357
0.66 39406.0323822039
0.665 39402.4105105554
0.67 39398.7554221619
0.675 39395.0658506273
0.68 39391.0823609397
0.685 39387.3201505814
0.69 39383.2619523604
0.695 39379.164347408
0.7 39375.0257321928
0.705 39370.5883556926
0.71 39366.3629918216
0.715 39361.836093563
0.72 39357.0070450478
0.725 39352.383685622
0.73 39347.4548777124
0.735 39342.7263659329
0.74 39337.4358193141
0.745 39332.3406069654
0.75 39326.932919256
0.755 39321.462731684
0.76 39315.9270349082
0.765 39310.0724311262
0.77 39304.1468353437
0.775 39298.1466700548
0.78 39291.8196255168
0.785 39285.411327947
0.79 39278.6701839641
0.795 39271.8401524324
0.8 39264.9163871417
0.805 39257.6482874094
0.81 39250.0322744501
0.815 39242.3085567185
0.82 39234.470869298
0.825 39226.2697627804
0.83 39217.7001334012
0.835 39208.9977769737
0.84 39195.3483170907
0.845 39034.4989733901
0.85 38893.3008300722
0.855 38774.7693339119
0.86 38683.1331766262
0.865 38624.3171993069
0.87 38606.6502792055
0.875 38641.9171366321
0.88 38746.9442233095
0.885 38946.0291786895
0.89 39274.7273002784
0.895 39785.8647395735
0.9 40559.2852666378
0.905 41718.0064815411
};
\addlegendentry{\scriptsize UE}
\addplot[thick, blue!50.1960784313725!black]
table {%
0.55 40995.0465039258
0.555 40704.010125904
0.56 40433.5629362008
0.565 40182.5420405061
0.57 39949.9024182805
0.575 39734.7102652148
0.58 39536.137404263
0.585 39353.4567245741
0.59 39186.0386228219
0.595 39033.3484367723
0.6 38894.9448767592
0.605 38770.4794774662
0.61 38659.6971104355
0.615 38562.437617528
0.62 38478.6386477073
0.625 38408.3398046549
0.63 38351.6882416275
0.635 38308.9458735832
0.64 38280.4984160511
0.645 38266.8665068964
0.65 38266.8913653119
0.655 38269.2859695443
0.66 38271.98185649
0.665 38274.9805745197
0.67 38278.2822862468
0.675 38281.8881265768
0.68 38285.7978764845
0.685 38290.0118944448
0.69 38294.529654219
0.695 38299.3505072162
0.7 38304.4739179572
0.705 38309.8976425627
0.71 38315.6202885991
0.715 38321.639929822
0.72 38327.9540157782
0.725 38334.5584134882
0.73 38341.4504380806
0.735 38348.6246506981
0.74 38356.0771196611
0.745 38363.8018373321
0.75 38371.79244073
0.755 38380.0419348988
0.76 38388.5418884773
0.765 38397.2832151781
0.77 38406.2572082209
0.775 38415.4517557983
0.78 38424.8561823703
0.785 38434.455104078
0.79 38444.2364746068
0.795 38454.1822217618
0.8 38464.2769159276
0.805 38474.5002649513
0.81 38484.8306787109
0.815 38495.2479169995
0.82 38505.7248132729
0.825 38516.2364015507
0.83 38526.7532042599
0.835 38537.2413791401
0.84 38547.6693461355
0.845 38557.9969016601
0.85 38568.1854176474
0.855 38578.1896503395
0.86 38587.9612434302
0.865 38597.4476740852
0.87 38606.5927155518
0.875 38641.9171366321
0.88 38746.9442233095
0.885 38946.0291786895
0.89 39274.7273002784
0.895 39785.8647395735
0.9 40559.2852666378
0.905 41718.0064815411
};
\addlegendentry{\scriptsize SO}
\end{axis}
\end{tikzpicture}

%% file: betas.tikz
\begin{tikzpicture}

\definecolor{color0}{rgb}{0.75,0,0.75}

\begin{axis}[
width=4cm,
height=4cm,
legend cell align={left},
legend style={
  fill opacity=0,
  draw opacity=1,
  text opacity=1,
  at={(0.4,0.635)},
  anchor=south west,
  draw=black
},
legend cell align={left},
legend style={fill opacity=0.8, draw opacity=1, text opacity=1, draw=white!80!black},
tick align=outside,
tick pos=left,
x grid style={white!69.0196078431373!black},
xlabel={\scriptsize \(\displaystyle \alpha\)},
xmin=0.53225, xmax=0.92275,
xtick style={color=black},
y grid style={white!69.0196078431373!black},
title={\scriptsize $\beta$},
ymin=-0.0499985714285714, ymax=1.04997,
ytick style={color=black},
ylabel near ticks,
scaled y ticks = false,
ytick style={color=black},
ylabel near ticks,
scaled y ticks = false,
y tick label style={/pgf/number format/fixed, font = \scriptsize},
xlabel near ticks,
scaled x ticks = false,
x tick label style={/pgf/number format/fixed, font = \scriptsize},
]
\addplot [dashed, blue!50.1960784313725!black]
table {%
0.55 0.999971428571429
0.555 0.999971428571429
0.56 0.999971428571429
0.565 0.999971428571429
0.57 0.999971428571429
0.575 0.999971428571429
0.58 0.999971428571429
0.585 0.989
0.59 0.969457142857143
0.595 0.949914285714286
0.6 0.9304
0.605 0.910857142857143
0.61 0.891314285714286
0.615 0.8718
0.62 0.852285714285714
0.625 0.832742857142857
0.63 0.813228571428571
0.635 0.793714285714286
0.64 0.7742
0.645 0.754714285714286
0.65 0.7352
0.655 0.715714285714286
0.66 0.696228571428571
0.665 0.676742857142857
0.67 0.657257142857143
0.675 0.637771428571429
0.68 0.618314285714286
0.685 0.598828571428571
0.69 0.579371428571429
0.695 0.559914285714286
0.7 0.540457142857143
0.705 0.521028571428571
0.71 0.501571428571429
0.715 0.482142857142857
0.72 0.462742857142857
0.725 0.443314285714286
0.73 0.423914285714286
0.735 0.404485714285714
0.74 0.385114285714286
0.745 0.365714285714286
0.75 0.346342857142857
0.755 0.326971428571429
0.76 0.3076
0.765 0.288257142857143
0.77 0.268914285714286
0.775 0.249571428571429
0.78 0.230257142857143
0.785 0.210942857142857
0.79 0.191657142857143
0.795 0.172371428571429
0.8 0.153085714285714
0.805 0.133828571428571
0.81 0.1146
0.815 0.0953714285714286
0.82 0.0761428571428571
0.825 0.0569428571428571
0.83 0.0377714285714286
0.835 0.0186
0.84 0
0.845 0
0.85 0
0.855 0
0.86 0
0.865 0
0.87 0
0.875 0
0.88 0
0.885 0
0.89 0
0.895 0
0.9 0
0.905 0
};
\addlegendentry{\scriptsize UE}
\addplot [thick, blue!50.1960784313725!black]
table {%
0.55 0.999971428571429
0.555 0.999971428571429
0.56 0.999971428571429
0.565 0.999971428571429
0.57 0.999971428571429
0.575 0.999971428571429
0.58 0.999971428571429
0.585 0.999971428571429
0.59 0.999971428571429
0.595 0.999971428571429
0.6 0.999971428571429
0.605 0.999971428571429
0.61 0.999971428571429
0.615 0.999971428571429
0.62 0.999971428571429
0.625 0.999971428571429
0.63 0.999971428571429
0.635 0.999971428571429
0.64 0.999971428571429
0.645 0.999971428571429
0.65 0.989771428571429
0.655 0.968628571428571
0.66 0.947514285714286
0.665 0.926228571428571
0.67 0.905
0.675 0.883571428571429
0.68 0.8622
0.685 0.840628571428571
0.69 0.819142857142857
0.695 0.797428571428571
0.7 0.775828571428571
0.705 0.754
0.71 0.732
0.715 0.710057142857143
0.72 0.687942857142857
0.725 0.665914285714286
0.73 0.643714285714286
0.735 0.621628571428571
0.74 0.599371428571429
0.745 0.576942857142857
0.75 0.554657142857143
0.755 0.532228571428571
0.76 0.509685714285714
0.765 0.486971428571429
0.77 0.464171428571429
0.775 0.4416
0.78 0.4186
0.785 0.395885714285714
0.79 0.373142857142857
0.795 0.350028571428571
0.8 0.326914285714286
0.805 0.303828571428571
0.81 0.280857142857143
0.815 0.2576
0.82 0.234514285714286
0.825 0.211228571428571
0.83 0.1882
0.835 0.164685714285714
0.84 0.141542857142857
0.845 0.118028571428571
0.85 0.0946
0.855 0.0714
0.86 0.048
0.865 0.0245428571428571
0.87 0.00111428571428571
0.875 0
0.88 0
0.885 0
0.89 0
0.895 0
0.9 0
0.905 0
};
\addlegendentry{\scriptsize SO}
\end{axis}
\end{tikzpicture}

%% file: fares.tikz
\begin{tikzpicture}

\definecolor{color0}{rgb}{0.12156862745098,0.466666666666667,0.705882352941177}

\begin{axis}[
width=6.5cm,
height=3.5cm,
tick align=outside,
tick pos=left,
x grid style={white!69.0196078431373!black},
xlabel={\scriptsize \(\displaystyle \alpha\)},
xmin=0.53225, xmax=0.92275,
xtick style={color=black},
y grid style={white!69.0196078431373!black},
ylabel={\scriptsize $\tau_p$},
ymin=-0.266255153296966, ymax=0.0126788168236651,
ytick style={color=black},
y tick label style={/pgf/number format/fixed, font = \scriptsize},
xlabel near ticks,
scaled x ticks = false,
x tick label style={/pgf/number format/fixed, font = \scriptsize},
]
\addplot [thick, blue!50.1960784313725!black]
table {%
0.585 -0.253576336473301
0.59 -0.234584758802208
0.595 -0.216124202739534
0.6 -0.198102401093545
0.605 -0.180427091810507
0.61 -0.16300513570276
0.615 -0.145741582417488
0.62 -0.128538668668244
0.625 -0.111294730667676
0.63 -0.0939030101229961
0.635 -0.0762503299903725
0.64 -0.0582156123131649
0.645 -0.0396682057450519
0.65 -0.0297096468677949
0.655 -0.0294833767382711
0.66 -0.0292662386153257
0.665 -0.0290315632514068
0.67 -0.0288060979856072
0.675 -0.0285624719578973
0.68 -0.028328279310167
0.685 -0.0280754030479805
0.69 -0.0278323674050501
0.695 -0.0275702641021281
0.7 -0.0273186425624721
0.705 -0.0270477581644875
0.71 -0.0267570687203207
0.715 -0.0264778385665561
0.72 -0.0261789494938808
0.725 -0.0258929953257516
0.73 -0.0255879108227553
0.735 -0.0252977804049031
0.74 -0.0249895581855696
0.745 -0.0246632829959482
0.75 -0.0243555970112211
0.755 -0.0240318892797548
0.76 -0.0236927112952379
0.765 -0.0233388073516744
0.77 -0.0229711482838216
0.775 -0.022631873803613
0.78 -0.0222417478895492
0.785 -0.0218856162058211
0.79 -0.0215250929884489
0.795 -0.0211179070948272
0.8 -0.0207105082938238
0.805 -0.0203070555121087
0.81 -0.019912531742162
0.815 -0.0194821028876833
0.82 -0.0190705074612351
0.825 -0.0186318721004463
0.83 -0.0182264309527498
0.835 -0.0177500044837127
0.84 -0.0173234402033435
0.845 -0.0168381547926208
0.85 -0.0163628236582734
0.855 -0.0159110123420072
0.86 -0.015430019357392
0.865 -0.0149318675326641
0.87 -0.0144320685331081
};
\addplot [dashed, black]
table {%
0.585 -0.266255153296966
0.585 0.0126788168236651
};
\addplot [dashed, black]
table {%
0.87 -0.266255153296966
0.87 0.0126788168236651
};
\addplot [thick, blue!50.1960784313725!black]
table {%
0.55 0
0.555 0
0.56 0
0.565 0
0.57 0
0.575 0
0.58 0
0.585 0
};

\end{axis}

\end{tikzpicture}